\documentclass[10pt]{article}

\usepackage{amsmath,amssymb,epsfig,graphics}

\def\etal{\emph{et al} }

\def\be{\begin{equation}}
\def\ee{\end{equation}}
\def\lb{\label}

\def\d{{\rm d}}
\def\e{{\rm e}}
\def\s{{\rm sech}(w\tau)}
\def\t{{\rm tanh}(w\tau)}

\def\sN{\mathcal{N}}
\def\EEE{E_1{}^1}
\def\Sm{\Sigma_-}
\def\Sc{\Sigma_\times}
\def\Nm{N_-}
\def\Nc{N_\times}
\def\Sp{\Sigma_+}
\def\Np{N_+}

\def\hSm{\hat{\Sigma}_-}
\def\hSc{\hat{\Sigma}_\times}
\def\hNm{\hat{N}_-}
\def\hNc{\hat{N}_\times}

\def\Smh{{\Sigma_-^H}}
\def\Sch{{\Sigma_\times^H}}
\def\Nmh{{N_-^H}}
\def\Nch{{N_\times^H}}
\def\Sph{{\Sigma_+^H}}

\def\di{2e^{-\tau}\tfrac{\partial}{\partial x}}
\def\E{\mathcal{E}}
\def\H{\mathcal{H}}
\def\Em{\E_-}
\def\Ec{\E_\times}
\def\Hm{\H_-}
\def\Hc{\H_\times}
\def\Ep{\E_+}
\def\Hp{\H_+}

\def\prd{{\it Phys. Rev. D} }
\def\prl{{\it Phys. Rev. Lett.} }
\def\cqg{{\it Class. Quantum Grav.} }

\begin{document}

\begin{center}
{\large\bf

New explicit spike solution -- non-local component of the generalized 
Mixmaster attractor}

\

Woei Chet Lim

Department of Physics, Princeton University, Princeton, NJ 08544,
USA.

Email: wlim@princeton.edu

\

\today
\end{center}

\begin{abstract}
By applying a standard solution-generating transformation to an arbitrary 
vacuum Bianchi type II solution, one generates a new solution with spikes 
commonly observed in numerical simulations. It is conjectured that the 
spike solutions are part of the generalized Mixmaster attractor.
\end{abstract}

\section{Introduction}

Berger and Moncrief \cite{BM93} studied Gowdy spacetimes and
found small-scale spatial structures develop on approach to the initial 
singularity. Since then many efforts have been spent trying to understand 
these spiky structures in Gowdy spacetimes and in more general $G_2$ 
spacetimes through numerical simulations and analytical approximations
\cite{BG98,HS98,RW01,GW03,AELU05,thesis}. 

My motivation in studying spikes is to understand its role on
approach to generic singularities.
Lifshitz, Khalatnikov and Belinskii \cite{LK63,BKL70,BKL82}
were the first to provide heuristic arguments that the approach 
to generic spacelike singularities are
\emph{vacuum dominated, local, and oscillatory} (known as the BKL 
conjecture).
Further evidence came from the study of Bianchi type IX cosmologies
by Misner \cite{M69a,M69b,MTW}, who coined the term ``Mixmaster" to 
describe the oscillatory behaviour.
Uggla \etal \cite{UEWE03} provided a detailed description of the 
local attractor for generic singularities (called the \emph{generalized 
Mixmaster attractor}).
See \cite{HUR07} for a more complete introduction and the latest work on 
the attractor (called the \emph{billiard attractor}).
The \emph{local} part of the BKL conjecture is increasingly under 
challenge from numerical evidence of the presence of 
\emph{recurring transient} spikes, which are non-local structures, in the 
approach to singularities \cite{AELU05,thesis}.
Insufficiently resolved spiky structures in the singular regime have been
observed in numerical simulations of $U(1)$ symmetric spacetimes
\cite{BM98},\cite[Chapter 7]{H99} and generic spacetimes \cite{G04}.
 
In this paper I present the first explicit spike solutions in the class of
Gowdy spacetimes. I then conjecture that spikes are an integral part 
of the dynamics on approach to generic singularities, and should 
constitute the non-local part of the generalized Mixmaster attractor.

\section{Gowdy spacetimes}

Gowdy spacetimes refer to solutions of the vacuum Einstein equations, 
with metric of the form
\be
ds^2 = - \e^{(\lambda-3\tau)/2} \d\tau^2 + \e^{(\lambda+\tau)/2} \d x^2
	+ \e^{P-\tau}(\d y+Q\, \d z)^2 + \e^{-P-\tau} \d z^2.
\ee
The vacuum Einstein equations imply that $P(\tau,x)$ and $Q(\tau,x)$ are 
determined up to a constant by
\begin{align}
\lb{Ptt}
	P_{\tau\tau} &= \e^{2P} Q_\tau^2 + \e^{-2\tau}(P_{xx}-\e^{2P}Q_x^2)
\\
\lb{Qtt}
	Q_{\tau\tau} &= -2 P_\tau Q_\tau + \e^{-2\tau}(Q_{xx} + 2 P_x Q_x),
\end{align}
while $\lambda(\tau,x)$ decouples from the above equations and is 
determined up to a constant by
\begin{align}
	\lambda_\tau &= -(P_\tau^2 + \e^{2P} 
			Q_\tau^2)-\e^{-2\tau}(P_x^2+\e^{2P} Q_x^2)
\\
	\lambda_x &= -2(P_\tau P_x + \e^{2P} Q_\tau Q_x).
\end{align}
The decoupling of $\lambda$ simplifies the analysis of Gowdy models, as 
focus can be narrowed down to $P,Q$.
The time variable $\tau$ tends to infinity as the singularity is 
approached.
The so-called $G_2$ spacetimes are those which admits two commuting 
Killing vector fields acting in the $(y,z)$ planes (or cylinders or tori), 
and Gowdy spacetimes are a special case, in which the action of the 
$G_2$ group is orthogonally transitive.

The dynamics of Gowdy spacetimes is of interest as part of the larger 
set of oscillatory behaviour, because it describes the dynamics during a 
Kasner era (see e.g. \cite[Section 7]{HUR07}). Although the Kasner era in 
Gowdy spacetimes terminates at Kasner solutions, in the more general $G_2$ 
models a Kasner era is followed by another Kasner era, without 
termination, hence giving the oscillatory Mixmaster dynamics.

\section{Solution-generating transformation}

The solution-generating transformation in interest is essentially the one 
used in \cite{RW01}. It is composed of two 
transformations. The first is modified inversion in the $(P,Q)$ 
hyperbolic plane \cite{RW01},\cite[eq (3.11)]{BM93}:
\be
\label{rot_trans}
	\e^{-\hat{P}} = \frac{e^{-P}}{Q^2+\e^{-2P}},\qquad
	\hat{Q} = -\frac{Q}{Q^2+\e^{-2P}},
\ee
where I have put a minus sign in $\hat{Q}$ to make it a reflected 
inversion, so that it has a better interpretation as a frame rotation below.
$\lambda$ is unaffected.
The second is the solution-generation transformation called the 
Gowdy-to-Ernst transformation \cite{RW01}:
\be
\label{GE_trans}
	\hat{P} = -P +\tau,\qquad
	\hat{Q}_{\tau}=-\e^{2(P-\tau)}Q_{x},\qquad
	\hat{Q}_{x}=-\e^{2P}Q_{\tau}.
\ee
The obstacle in generating explicit solutions is the integration to obtain 
$\hat{Q}$.
Composing the two transformations gives a powerful solution-generating 
transformation that can be iterated to generate a family of new 
solutions.

The transformation and the solutions can be presented more elegantly in
the orthonormal frame formulation with the so-called $\beta$-normalized 
variables as presented in \cite[Section 4.4.1]{EUW02}, in 
the so-called timelike area gauge with 
\be
\lb{NE}
	t=\tau,\qquad \sN_0^{-1}=-2,\qquad \EEE=2\e^{-\tau}.
\ee 
In the orthonormal frame formulation, an orthonormal 
frame is used, and variables are the frame components and (essentially) 
their first derivatives, divided by $\beta$.
The $(y,z)$ area expansion rate $\beta$ is related to $\lambda$ by
\cite[App A.3]{EUW02}
\be
	\beta=-\tfrac12\e^{-(\lambda-3\tau)/4}.
\ee
Negative value for $\beta$ describes contraction as $\tau$ increases.
 
For Gowdy models, the key $\beta$-normalized variables are 
$(\Sm,\Nc,\Sc,\Nm)$, which are decompositions of
orthonormal frame components of the 3-by-3 $\Sigma_{\alpha\beta}$ and 
$N_{\alpha\beta}$ matrices:
\begin{align}
\lb{matrix}
        \Sigma_{\alpha \beta}
        &= \left(\begin{matrix}
        -2\Sp & 0 & 0 \\
        0 & \Sp + \sqrt{3} \Sm & \sqrt{3} \Sc \\
        0 & \sqrt{3} \Sc & \Sp - \sqrt{3} \Sm
        \end{matrix}\right)
\\
\lb{matrix2}
        N_{\alpha \beta}
        &= \left(\begin{matrix}
        0 & 0 & 0 \\
        0 & 2\sqrt{3} \Nm & \sqrt{3} \Nc \\
        0 & \sqrt{3} \Nc & 0
        \end{matrix}\right),
\end{align}
where $\Sp$ is given by
\be
        \Sp = \tfrac12(1-\Sm^2-\Sc^2-\Nm^2-\Nc^2).
\ee
In order to preserve the form of (\ref{matrix2}),
the spatial orthonormal frame itself is rotating around the $x$-axis at 
the rate of $R=-\sqrt{3}\Sc$ \cite[Section 3.1]{EUW02}.
Under this condition, $(\Sm,\Nc,\Sc,\Nm)$ are related to
the derivatives of $P,Q$ by
\be
\Sm = -\frac{P_\tau}{\sqrt{3}},\qquad
\Nc = -\frac{\e^{-\tau}P_x}{\sqrt{3}},\qquad
\Sc = -\frac{\e^P Q_\tau}{\sqrt{3}},\qquad
\Nm =  \frac{\e^{P-\tau} Q_x}{\sqrt{3}}.
\ee

The reflected inversion transformation has a simple interpretation as 
the
rotation of the spatial orthonormal frame to another rotation rate that 
also preserves the form of (\ref{matrix2}). The angle of rotation $\phi$ 
is determined by
\be
\lb{rotation}
	\cos 2\phi = \frac{(Q\e^P)^2-1}{(Q\e^P)^2+1},\qquad
	\sin 2\phi = \frac{2Q\e^P}{(Q\e^P)^2+1}.
\ee
$(\Sm,\Sc)$ rotates as follows:
\be
	\left(\begin{matrix}
	\hSm \\
	\hSc \end{matrix}\right)
	= \left(\begin{matrix}
	\cos 2\phi & \sin 2\phi \\
	- \sin 2\phi & \cos 2\phi \end{matrix}\right)
	\left(\begin{matrix}
        \Sm \\
        \Sc \end{matrix}\right),
\ee
and similarly for $(\Nm,\Nc)$.

The Gowdy-to-Ernst transformation is much simpler in terms of
$(\Sm,\Nc,\Sc,\Nm)$:
\be 
\label{GE_map}
	(\hSm,\hNc,\hSc,\hNm) = (-\Sm-\tfrac{1}{\sqrt{3}},-\Nc,\Nm,\Sc).
\ee

\section{The explicit solutions}

I now present the Kasner solutions, 
the rotated Kasner solutions,
the Taub vacuum solutions,
the rotated taub solutions, 
and the spike solutions. 
I shall discuss their dynamics in the next section.

\subsection{The Kasner solutions}

We shall use the Kasner solutions (vacuum Bianchi type I solutions) as the 
seed solution. The metric components $(P,Q,\lambda)$ are given by
\be
	P = w \tau + P_0,\qquad
	Q = Q_0,\qquad
	\lambda = - w^2 \tau + \lambda_0,
\ee
with $w$ serving to parametrize the Kasner solutions.
$w$ is related to the Khalatnikov-Lifshitz parameter $u$ by 
$w=2u+1$.
$P_0$, $Q_0$ and $\lambda_0$ are arbitrary constants and do not 
parametrize the Kasner solutions.

In terms of $\beta$-normalized variables, the Kasner solutions are given 
by
\be
\label{Kas_beta}
	(\Sm,\Nc,\Sc,\Nm) = (-\frac{w}{\sqrt{3}},0,0,0).
\ee
The flat Kasner solutions are those with $w=\pm1, \infty$, and the 
other plane-symmetric Kasner solutions are those with $w=0,\pm3$.

\subsection{The rotated Kasner solutions}

It is essential to choose $Q_0\neq0$ so that the rotation transformation 
is nontrivial.
Applying the rotation transformation (\ref{rot_trans}) to the Kasner 
solutions yields the Kasner solutions viewed in a rotating frame (see 
also \cite[eqs (3.12)--(3.13)]{BM93}, \cite[eq (7)]{BG98}). 
The metric components $(P,Q,\lambda)$ are given by
\begin{align}
	\e^{-P} &= \frac{\e^{-(w\tau+P_0)}}{Q_0^2+\e^{-2(w\tau+P_0)}}
\\
	Q &= -\frac{Q_0}{Q_0^2+\e^{-2(w\tau+P_0)}}
\\
	\lambda &= - w^2 \tau + \lambda_0.
\end{align} 
It turns out to be convenient to write $P$ and $Q$ in the form
\be
	\e^{-P} = \frac{\text{sech}(w\tau+P_0+\ln Q_0)}{2Q_0},\qquad
	Q = -\frac{1}{2Q_0} [1+\text{tanh}(w\tau+P_0+\ln Q_0)].
\ee
For simplicity we will now choose $P_0+\ln Q_0=0$.
The $\beta$-normalized variables are then given by
\be
\label{rot_beta}
        (\Sm,\Nc,\Sc,\Nm) =
        \left(  
  -\frac{w}{\sqrt{3}}\t, 0,
   \frac{w}{\sqrt{3}}\s,0\right).
\ee

\subsection{The Taub vacuum solutions}

Applying the composed transformation 
(\ref{rot_trans})--(\ref{GE_trans}) on the Kasner solutions with $w\neq0$
(For the $w=0$ Kasner, the transformation gives the $w=1$ Kasner.)
yields the Taub vacuum solutions (vacuum Bianchi type II solutions), whose 
$(P,Q,\lambda)$ are given by
\begin{align}
	P &= \tau + \ln(\s) - \ln(2Q_0)
\\
	Q &= 2 Q_0 w  x + Q_1 
\\
	\lambda &= -2 \ln(\s) - (w^2+1)\tau + \lambda_1.
\end{align}
$Q_1$ and $\lambda_1$ are arbitrary constants and do 
not parametrize the Taub solutions.
The $\beta$-normalized variables are given by
\be
	(\Sm,\Nc,\Sc,\Nm) = 
	\left(  \frac{w}{\sqrt{3}}\t-\frac{1}{\sqrt{3}}, 0, 0,
		\frac{w}{\sqrt{3}}\s \right).
\ee

\subsection{The rotated Taub solutions (the false spike solutions)}

Note that the frame of the Taub solutions is not rotating.
Applying the rotation transformation (\ref{rot_trans}) to the Taub 
solutions yields the rotated Taub solutions:
\begin{align}
        P &= -\tau 
	- \ln(\s)-\ln[(w\e^{\tau}\s x)^2+1] +\ln(2Q_0)
\\
	Q &= -\frac{1}{2Q_0} \frac{wx (\e^\tau\s)^2}{(w\e^{\tau}\s x)^2+1}
\\
        \lambda &= -2 \ln(\s) - (w^2+1)\tau + \lambda_1,
\end{align}
where for simplicity we have chosen $Q_1=0$, so that the term
$x-Q_1/(2wQ_0)$ simplifies to $x$.
The $\beta$-normalized variables are given by
\be
\lb{false_spike}
        (\Sm,\Nc,\Sc,\Nm) =
        \left(c \Sm{}_\text{Taub},
        -s\Nm{}_\text{Taub},
        -s \Sm{}_\text{Taub},
        c \Nm{}_\text{Taub}
        \right),
\ee
where
\be
\lb{csf}
        c = \frac{f^2-1}{f^2+1},\qquad
        s = \frac{2f}{f^2+1},\qquad
        f = (Q\e^P)_\text{Taub}=w \e^\tau \s x.
\ee
The rotation transformation makes the frame rotates in a spiky manner near 
$x=0$. The rotated Taub solutions are also referred to as the false spike 
solutions.
The shape of the spike will be described in the next subsection.

\subsection{The spike solutions}

Applying the composed transformation 
(\ref{rot_trans})--(\ref{GE_trans})
on the Taub vacuum solutions yields
a new solution, whose $(P,Q,\lambda)$
are given by
\begin{align}
	P &= 2\tau + \ln(\s)-\ln[(w\e^{\tau}\s x)^2+1] -\ln(2Q_0)
\\
	Q &= -4Q_0 w [\e^{-2\tau} + 2(w\t-1)x^2] + Q_2
\\
	\lambda &=-4\ln(\s) + 2\ln[(w\e^{\tau}\s x)^2+1] -(w^2+4)\tau 
		+\lambda_2.
\end{align}
The $\beta$-normalized variables are given by
\be
\lb{spike}
        (\Sm,\Nc,\Sc,\Nm) =
	\left(-c \Sm{}_\text{Taub} -\frac{1}{\sqrt{3}},
	s\Nm{}_\text{Taub},
	c \Nm{}_\text{Taub},
	-s \Sm{}_\text{Taub}
	\right).
\ee
The spatial dependence of $(\Sm,\Nc,\Sc,\Nm)$ lies in
$c$ and $s$, which depends on $x$ in a spiky way.
$c$ has a spike of the form 
\be
	\frac{(kx)^2-1}{(kx)^2+1},
\ee
which has a single peak at $x=0$ and two zeros at $x=\pm 1/k$.
$s$ has a spike of the form
\be
	\frac{2kx}{(kx)^2+1},
\ee
which has two peaks with opposite signs at $x=\pm 1/k$ and a zero at 
$x=0$.
From (\ref{csf}), we have $k=w \e^\tau \s$, so the location of the peaks 
varies with time. 
The limit of $k$ as $\tau \rightarrow \infty$ bifurcates at $|w|=1$, as
can be seen clearly when $k$ is written in the form
\be
\label{k_form}
	k= w \e^{\tau-|w\tau|}(1+\tanh|w\tau|).
\ee
For $|w|>1$, $k$ is largest (and the spike is narrowest) when 
\be
\lb{t_narrow}
	\tau=\frac1w \text{arctanh} \frac1w,
\ee
and $k$ tends to zero as $\tau \rightarrow \infty$.
For $0<|w|<1$, $k$ tends to infinity (and the spike increasingly narrow) 
as 
$\tau \rightarrow \infty$.

\subsection{The radius of the spike}

Therefore, we shall define the radius of the spike to be one half the 
$x$-coordinate distance between the two peaks in $s$, which works out to 
be
\be
\lb{x_rad}
        \text{radius of spike} = \frac{1}{w} \e^{-\tau}\cosh w\tau.
\ee
We must measure the radius of the spike against another important scale -- 
the particle horizon (when approaching the initial singularity) or the 
event horizon (when approaching the final singularity),
whose $x$-coordinate radius is given by
\be
        \text{radius of horizon} = \int_\tau^\infty |\sN_0| \EEE d\tau =
\e^{-\tau},
\ee
as follows from (\ref{NE}).
Thus, the radius of the spike as a multiple of the horizon is
\be
\lb{spike_rad}
        \text{radius of spike} = \frac{1}{w}\cosh w\tau \times
        \text{radius of horizon}.
\ee
This measure of the radius is more meaningful than (\ref{x_rad}).
Using (\ref{spike_rad}), the spike is narrowest relative to the 
horizon at $\tau=0$, which does not coincide with (\ref{t_narrow}).

Notice that for $0<|w|<1$, the size of the spike is always
\emph{super-horizon}, despite appearing to be increasingly narrow.
On the other hand, for $|w|>1$, there
is a brief period during which the spike spike becomes a 
\emph{sub-horizon} structure. Physically, during this period, an observer 
along the \emph{spike worldline} $x=0$
would see substantial inhomogeneity. Mathematically, the
second-order spatial derivative terms $P_{xx}$ and $Q_{xx}$ in
the Einstein equations are not negligible during this period.

\begin{figure}[t!]
  \begin{center}
    \resizebox{\linewidth}{!}{\includegraphics{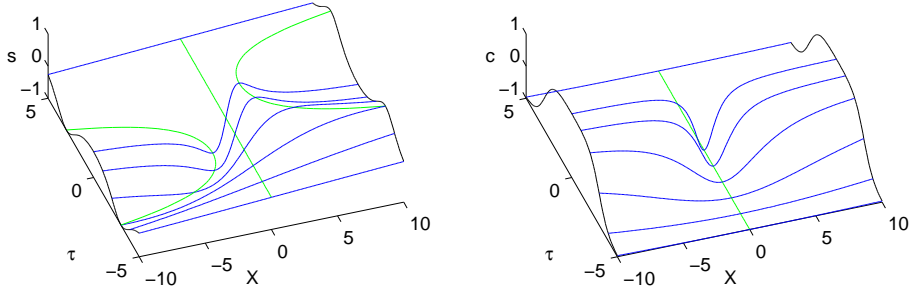}}
    \caption{The spiky structures of the $s$ and $c$ functions in 
(\ref{csf}) with $w=1$ against $\tau$ and $X=\e^\tau x$.}
    \label{fig_cs}
\end{center}
\end{figure}

To see the sub-horizon structures, it is useful to plot variables against
multiples of the radius of horizon, i.e. against the variable
\be
	X=\e^\tau x, 
\ee
in which the horizon of the observer along $x=0$ is
represented by $X=\pm1$. Figure~\ref{fig_cs} plots the spiky structures of
the $s$ and $c$ functions in (\ref{csf}) against $\tau$ and $X$.

\begin{figure}[t!]
  \begin{center}
    \resizebox{6cm}{!}{\includegraphics{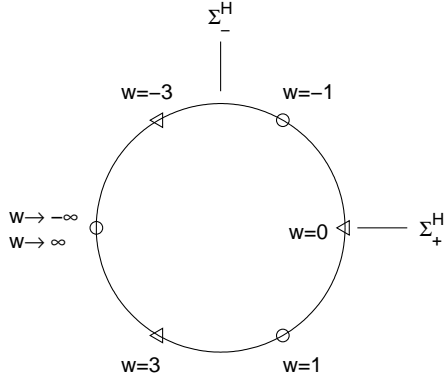}}
    \caption{The Kasner circle and the parameter $w$.}
    \label{fig_kas}
\end{center}
\end{figure}

\section{Visualizing the dynamics of the spike solutions}

To visualize the dynamics of the spike solutions, we will use both 
$\beta$- and Hubble-normalized variables.
For the spatially homogeneous background dynamics, it is best to use the 
Hubble-normalized variables (see \cite[Chapter 6]{WE97}),
which are related to the $\beta$-normalized ones via
\be
	(\Sp,\Sm,\Sc,\Nm,\Nc)^H = \frac{1}{1-\Sp}(\Sp,\Sm,\Sc,\Nm,\Nc),
\ee
and satisfy
\be
	\Sph {}^2+\Smh {}^2+\Sch {}^2+\Nmh {}^2+\Nch {}^2=1.
\ee
In the state space of Hubble-normalized variables, the Kasner solutions 
appear as \emph{equilibrium points} on a unit circle in the $(\Sp,\Sm)^H$ 
plane 
(see Figure \ref{fig_kas}). The three different representations of the flat 
Kasner solution are marked by a circle, and the three representations of 
the other axis-symmetric Kasner solution are marked by a triangle.
All other Kasner solutions have six different representations, each 
located equidistant from a flat Kasner point.

\begin{figure}[t!]
  \begin{center}
    \resizebox{8cm}{!}{\includegraphics{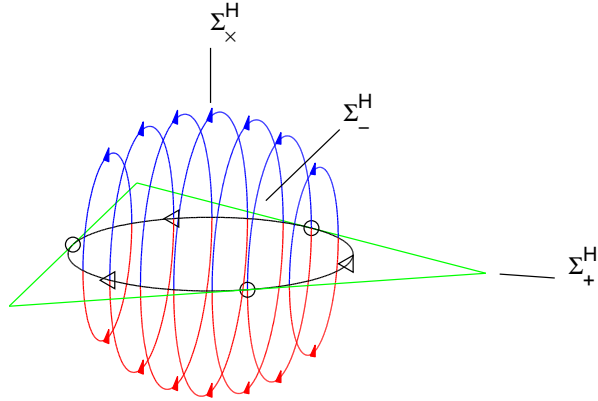}}
    \caption{The rotation orbits (frame transition set) on the sphere 
$\Sph{}^2+\Smh{}^2+\Sch{}^2=1$. Arrows indicate the direction of 
increasing $\tau$, towards the singularity.}
    \label{fig_rot}
\end{center}
\end{figure}

\begin{figure}[t!]
  \begin{center}
    \resizebox{8cm}{!}{\includegraphics{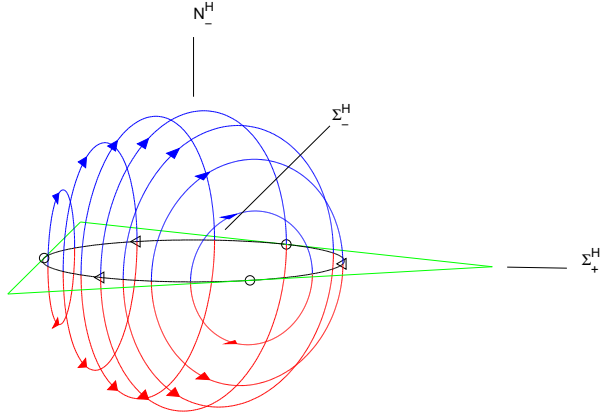}}
    \caption{The Taub orbits (curvature transition set) on the sphere 
$\Sph{}^2+\Smh{}^2+\Nmh{}^2=1$.}
    \label{fig_taub}
\end{center}
\end{figure}

\begin{figure}[t!]
  \begin{center}
    \resizebox{\linewidth}{!}{\includegraphics{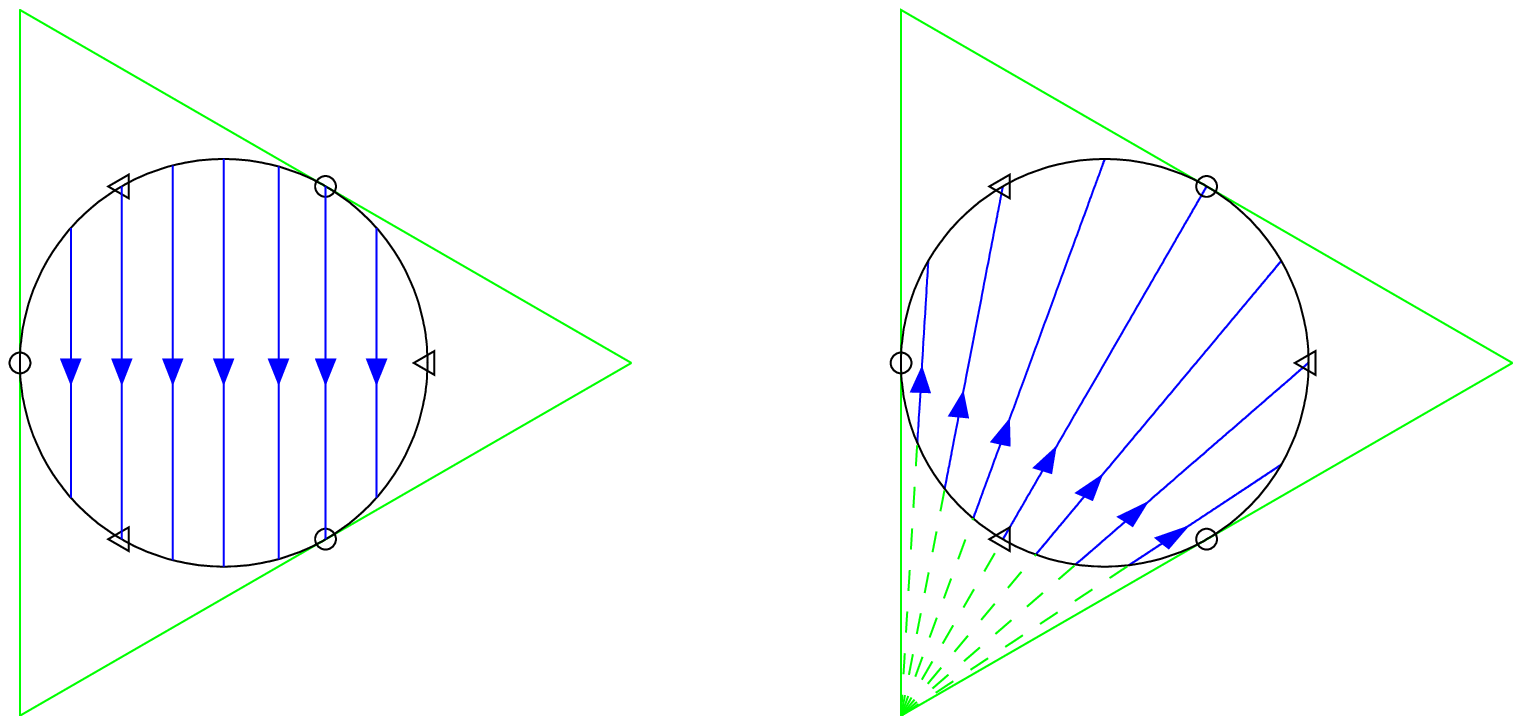}}
    \caption{The rotation and Taub orbits projected on the $(\Sp,\Sm)^H$ 
plane.}
    \label{fig_top}
\end{center}
\end{figure}

Each $w$-rotated Kasner solution appears as a semi-circular orbit on 
the sphere 
$\Sph{}^2+\Smh{}^2+\Sch{}^2=1$ (Figure \ref{fig_rot}), connecting
two representations of the same Kasner solution.
As $\tau$ goes from $-\infty$ to $\infty$,
the $w$-rotated Kasner solution tends from the $-|w|$-Kasner solution 
to the $|w|$-Kasner solution (also compare the limits of (\ref{rot_beta}) 
with (\ref{Kas_beta})).
The sign of $w$ determines the sign of $\Sch$.
The set of all such orbits (all values of $w\neq0$) is referred to as a
\emph{frame transition set} \cite{UEWE03}.

Each $w$-Taub solution appears as a semi-circular orbit on the sphere 
$\Sph{}^2+\Smh{}^2+\Nmh{}^2=1$ (Figure \ref{fig_taub}),
connecting two Kasner points.
As $\tau$ goes from $-\infty$ to $\infty$,
the $w$-Taub solution tends from the $(|w|+1)$-Kasner solution to 
the $(1-|w|)$-Kasner solution. 
The sign of $w$ determines the sign of $\Nmh$.
The set of all such orbits (all values of $w\neq0$) is referred to as a
\emph{curvature transition set}.

When projected on the $(\Sp,\Sm)^H$ plane, the frame transition sets form
parallel straight lines, while curvature 
transition sets form straight lines emanating from one corner of a
triangle superscribing the Kasner circle (Figure \ref{fig_top}).
The two transition sets combine to describe the dynamics during a 
Kasner era, which consists of long Kasner epochs (described by the Kasner 
equilibrium points), punctuated by brief periods of transitions.
During frame transitions, the rotating frame interrupts a Kasner epoch 
with a brief period of frame rotation,
thus fictitiously splits the same Kasner epoch into two epochs -- one with 
negative $w$ and one with positive $w$.
During curvature transitions, the spatial curvature becomes significant. 
Successive pairs of curvature and frame transitions reduce the value $|w|$ 
of the Kasner epoch by 2, and terminate at the final Kasner epoch with 
$0<w<1$. 
For examples,
\begin{align}
\label{eg1}
	w&=4.2
        \stackrel{\text{curv}}{\longrightarrow} -2.2
        \stackrel{\text{frame}}{\longrightarrow} 2.2
        \stackrel{\text{curv}}{\longrightarrow} -0.2
        \stackrel{\text{frame}}{\longrightarrow} 0.2,
\\
\label{eg2}
	w&=5.2
        \stackrel{\text{curv}}{\longrightarrow} -3.2 
        \stackrel{\text{frame}}{\longrightarrow} 3.2
        \stackrel{\text{curv}}{\longrightarrow} -1.2
        \stackrel{\text{frame}}{\longrightarrow} 1.2
        \stackrel{\text{curv}}{\longrightarrow} 0.8.
\end{align}
See \cite{HUR07} for a detailed discussion of these solutions.

Applying the rotation transformation to the Taub solutions yields the 
false spike solutions, which are simply the Taub solutions presented in 
a spatial frame that rotates in a spiky way. The Gowdy-to-Ernst 
transformation then maps the false spike solutions to the real ones.
The dynamics of the false spikes is similar to that of the real spikes 
(related through the simple map (\ref{GE_map})), so I shall focus on the 
real ones.

For the spike solution with value $w$,
as $\tau$ goes from $-\infty$ to $\infty$,
all orbits begin at the same Kasner point (with value $|w|+2$).
For $|w|\geq1$, the orbits also end at a common Kasner point (with 
value $2-|w|$). The smooth limit describes the \emph{transient spikes}.
For $0<|w|<1$, however, the orbit along $x=0$ ends at the Kasner point 
with value $2-|w|$, while orbits along $x\neq0$ ends at the Kasner point
with value $|w|$. This discontinuous limit describes the \emph{permanent 
spikes}. I shall discuss this in more detail in the next subsection.
The $|w|=1$ bifurcation of the limit is due to the limit of the factor $\s
\e^\tau$ as $\tau \rightarrow \infty$ (see eq (\ref{k_form})).

\begin{figure}[t!]
  \begin{center}
    \resizebox{\linewidth}{!}{\includegraphics{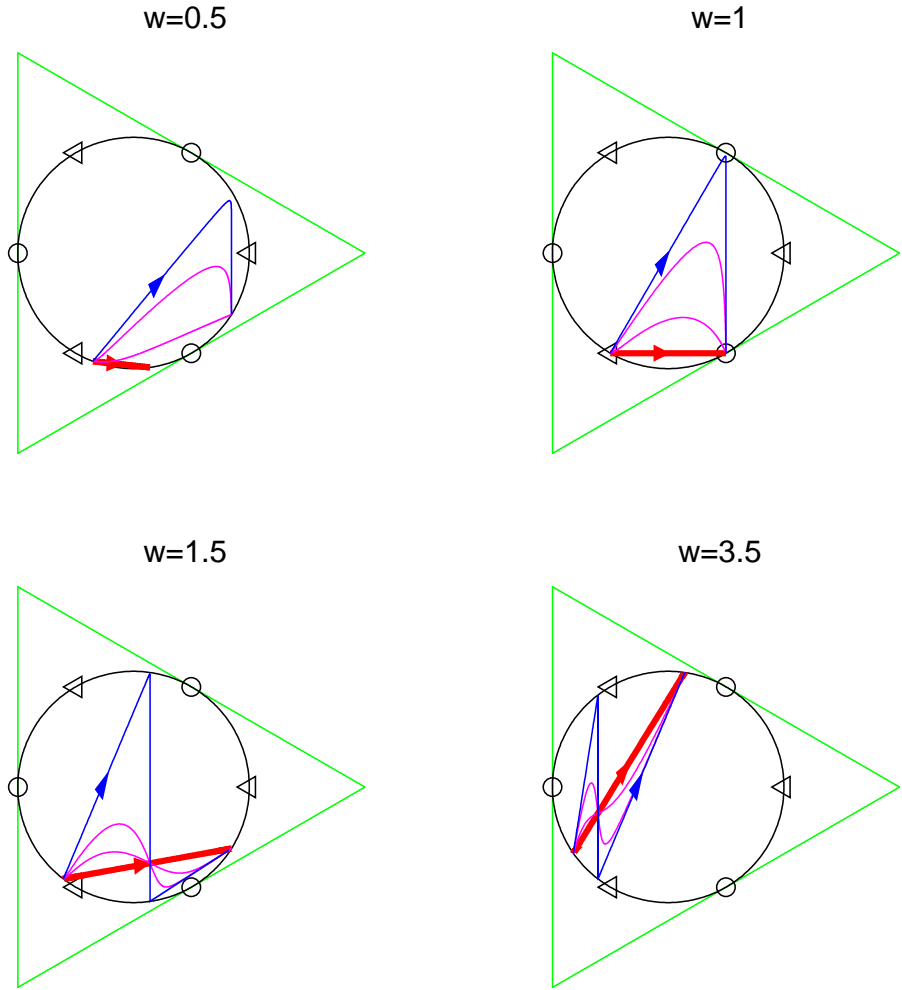}}
    \caption{Orbits of the spike solutions projected on the $(\Sp,\Sm)^H$ 
plane. Orbits are coloured red (thick arrowed line) along the spike 
worldline $x=0$, blue (thin arrowed line) along $x=1000$, and magenta 
(thin line without arrows) along small values of 
$x$. $w=0.5$, $1$, $1.5$, $3.5$ respectively.}
    \label{fig_spike}
\end{center}
\end{figure}

Along worldlines far away from the spike worldline, the 
orbits approximate the Taub and rotation orbits. Along the spike 
worldline, the orbit (called the \emph{spike orbit}) lies on the sphere 
$\Sph{}^2+\Smh{}^2+\Sch{}^2=1$, 
and forms a straight line when projected on the $(\Sp,\Sm)^H$ plane. 
Along worldlines near the spike worldlines, the orbits interpolate between 
the two extremes. 
The set of all orbits (all values of $w\neq0$) is called
the \emph{spike transition set}. Figure~\ref{fig_spike} shows four members
of the spike transition set. 

\begin{figure}[t!]
  \begin{center}
    \resizebox{\linewidth}{!}{\includegraphics{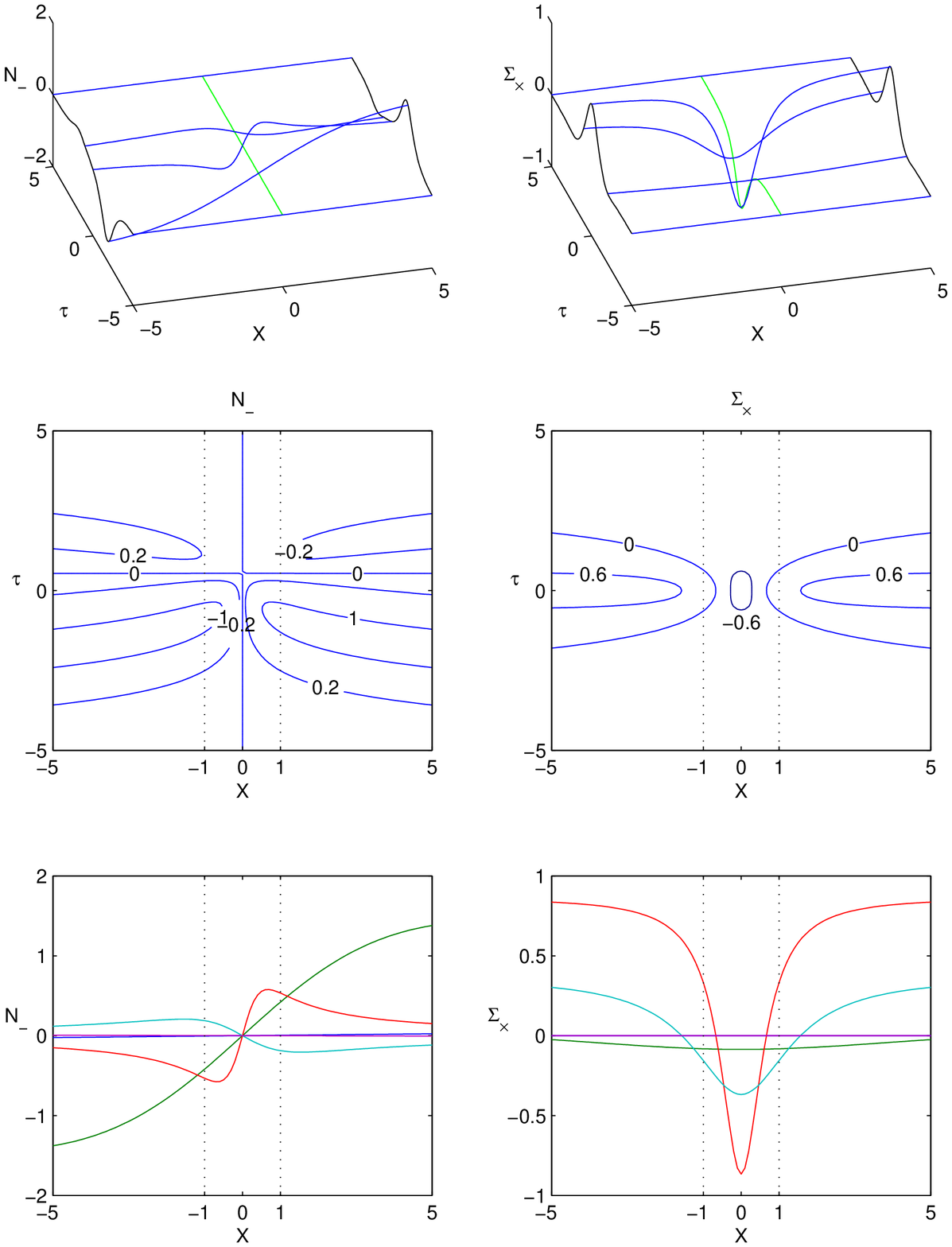}}
    \caption{Spikes in $\Nm$ and $\Sc$ with $w=1.5$.}
    \label{fig_NmSc}
\end{center}
\end{figure}

The inhomogeneous spike dynamics are best visualized in the
$\beta$-normalized $\Nm$ and $\Sc$, which are active variables in the 
Taub and rotation orbits respectively, and considered the active 
players in spike dynamics. 
Figure~\ref{fig_NmSc} shows the spikes in
$\beta$-normalized $\Nm$ and $\Sc$.

\subsection{Transient and permanent spikes}

We shall discuss the dynamics of transient spikes first.
For a spike solution with $|w|>1$ one gets a transient spike that smooths 
out as $\tau \rightarrow \infty$. 
In this case there is an instant in time
(which coincides with (\ref{t_narrow}))
\be
	\tau_\text{flip}=\frac1w \text{arctanh}\frac1w,
\ee
when the following variables become homogeneous:
\be
	\Nm(\tau_\text{flip},x)=0,\qquad 
	\Sm(\tau_\text{flip},x)=-\frac1{\sqrt{3}},\qquad 
	\Sp(\tau_\text{flip},x)=\frac{3-w^2}{6}. 
\ee
The spike in $\Nm$ flips sign at $\tau=\tau_\text{flip}$. All 
orbits also project on the same point on the $(\Sp,\Sm)^H$ plane at 
this instant in time. See Figures~\ref{fig_spike} and \ref{fig_NmSc}.
Dynamically, faraway observers undergo three transitions -- curvature, 
frame and curvature:
\be
	|w|+2 
	\stackrel{\text{curv}}{\longrightarrow} -|w| 
	\stackrel{\text{frame}}{\longrightarrow} |w|
	\stackrel{\text{curv}}{\longrightarrow} 2-|w|,
\ee
where the values $|w|+2$, $-|w|$, $|w|$ and $2-|w|$ are those of the 
consecutive Kasner epochs in the spike solution.
While faraway observers undergo the first curvature transition, the spike 
observer, because $\Nm(\tau,0)=0$, is unable to do so, and observes 
the formation of a spike, which starts as a super-horizon structure. 
While faraway observers undergo the frame transition,
the spike has narrowed to become a sub-horizon structure, reaches a 
minimum width at $\tau=0$, and then starts to widen, and flips sign at 
$\tau=\tau_\text{flip}$. The spike observer transitions directly to the 
fourth Kasner epoch:
\be
	|w|+2
	\longrightarrow 2-|w|.
\ee
While faraway observers undergo the second curvature transition,
the spike continues to widen and eventually smooths out.

We now discuss the formation of permanent spikes.
Recall that Gowdy models can undergo only one Kasner era. The 
$w$-value of the final Kasner epoch of the era satisfies $0 < w < 1$.
There are two different penultimate Kasner epochs -- one with 
$1 < w < 2$ and the other with $-1 < w < 0$. The former transitions to the 
final Kasner epoch through the curvature transition, and the latter 
through the frame transition. See examples (\ref{eg1})--(\ref{eg2}).

Permanent spikes form in the spike solutions with $0<|w|<1$.
Dynamically, faraway observers undergo two transitions -- curvature and 
frame, and terminate at the final Kasner epoch:
\be
        |w|+2
        \stackrel{\text{Taub}}{\longrightarrow} -|w|
        \stackrel{\text{rotation}}{\longrightarrow} |w|.
\ee
The spike observer on the other hand transitions to the 
penultimate Kasner epoch with $1<w<2$ as its final Kasner epoch:
\be
\label{spike_obs_trans}
	|w|+2
        \longrightarrow 2-|w|.
\ee
The resulting discontinuous pointwise (fixed $x$) limit is the permanent 
spike. 
One at first gets a transient spike that forms during the antepenultimate 
Kasner epoch with $2<w<3$, that later becomes a permanent spike after
the transition (\ref{spike_obs_trans}).
See Figure~\ref{fig_spike} with $w=0.5$.

Unfortunately, the spike solutions do not include permanent spikes 
that form directly during the penultimate Kasner epoch with $1<w<2$.
Spikes that form during the other penultimate Kasner epoch with $-1<w<0$ 
are false spikes.
Spikes do not form during the final Kasner epoch (with $0<w<1$).

\subsection{The Weyl invariants}

\begin{figure}[t!]
  \begin{center}
    \resizebox{\linewidth}{!}{\includegraphics{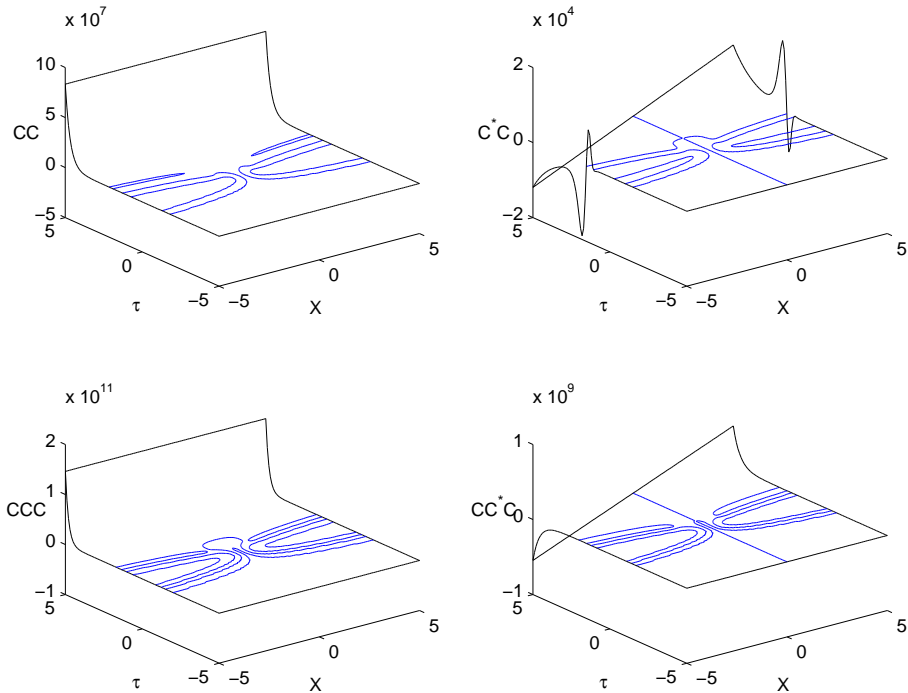}}
    \caption{The Weyl scalars for the spike solution with $w=2.5$.}
    \label{fig_weyl}
\end{center}
\end{figure}

For the spike solutions, the explicit expressions of the Weyl invariants
(formulae given in Appendix). 
are too complicated to be useful. Nonetheless we can compute them 
numerically to tell apart real and false spike solutions.
The Weyl scalars for a spike solution are shown in 
Figure~\ref{fig_weyl}. Observe the transient, sub-horizon structures in 
the Weyl scalars. 
The Weyl scalars blow up as $\tau \rightarrow\infty$, 
indicating the approach to a curvature singularity.
The Weyl scalars for the Kasner solutions, the rotated Kasner solutions, 
the Taub solutions and the false spike solutions are spatially 
homogeneous.

\begin{figure}[t!]
  \begin{center}
    \resizebox{\linewidth}{!}{\includegraphics{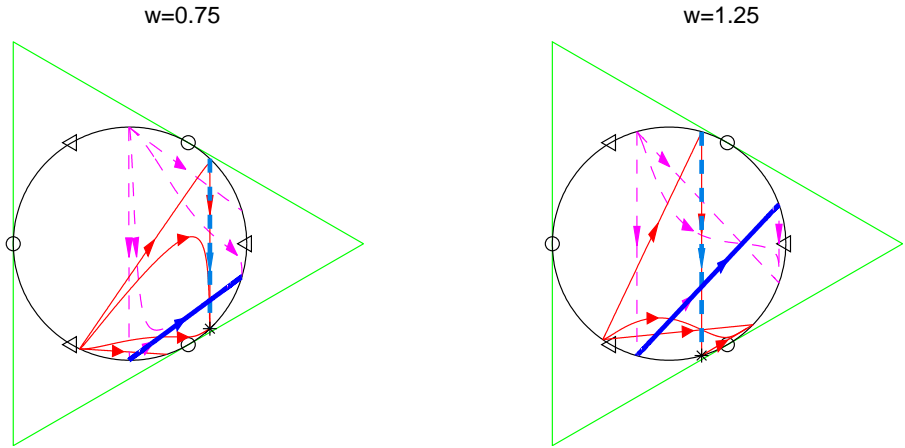}}
    \caption{Two families of orbits with $w=0.75$ and $1.25$. 
The Kasner seed is indicated by a black *, the rotation orbit in light 
blue (thick dashed line), the Taub orbit in dark blue (thick solid line), 
orbits of the false spike solution in magenta (thin dashed line), and 
orbits of the spike solution in red (thin solid line).}
    \label{fig_family}
\end{center}
\end{figure}

\subsection{Spike as the non-local part of the generalized Mixmaster 
attractor}

The spike solutions closely resemble those produced numerically
(see e.g. \cite{GW03, thesis}). This raises the question whether the 
spike solutions are an integral part of the generalized Mixmaster 
attractor.

Applying the rotation transformation and the Gowdy-to-Ernst transformation 
to the Kasner solutions
successively generates solutions describing the frame transition, the 
curvature transition, the false spike transition, and the (true) spike 
transition. See Figure~\ref{fig_family} for two families of orbits 
generated.
The Kasner circle, the frame transition set and the curvature transition 
set are known to be part of the generalized Mixmaster attractor 
(see e.g. \cite[Section IV]{UEWE03},\cite{HUR07}).
Thus it is natural to expect that the transformations preserve this 
property, and extend it to the false and true spike solutions, provided 
that spikes do not disappear. 
This is supported by numerical evidence, 
which indicates that spikes generally recur on the same spot (where 
$\Nm=0$), despite numerical evidence that pairs of zeros of $\Nm$ can be 
created or annihilated.
See \cite[concluding remarks]{HUR07} for a discussion of interaction
between spikes.

I therefore conjecture that the spike transition set (as well as the false 
spike transition set) constitutes the non-local part of the 
generalized Mixmaster 
attractor. The \emph{local} part of the BKL conjecture, which states that 
the dynamics becomes asymptotically local as the singularity is 
approached, should be modified to allow for spikes, whose dynamics is 
non-local, but only occurs within the horizon of isolated worldlines.

\subsection{The simplest gauge for the spike solutions}

A natural question arises regarding the merit of using a rotating frame, 
which produces false spikes. It turns out that the spike solutions are not 
elementary in a non-rotating frame (because the time integral of $\Sc$ in 
(\ref{spike}) is not elementary). Furthermore, the net rotation in the 
spike solutions is non-zero and depends on $x$. As a result, the spike 
orbit does not appear as a straight line when projected on any plane, and 
all other orbits end on Kasner points lying on the $(\Sm,\Sc)^H$ plane 
with fixed $\Sph$. Therefore a simple description of the spike transition 
set in a rotating frame is preferable to a complicated one in a 
non-rotation frame.

We now consider other temporal gauges. The flipping of the spike occurs 
instantaneously in the timelike area gauge \cite{EUW02}, but not in other 
gauges. This singles out the timelike area gauge as the simplest temporal 
gauge to present and study the spike solutions.

\subsection{Higher-order spike solutions}
Applying the transformation on the spike solutions yields
another new solution, whose $(P,Q,\lambda)$ are given by
\begin{align}
        P &= 3\tau + \ln(\s)-\ln[(w \e^{\tau}\s x)^2+1]
                - \ln[f_2^2+1]      -\ln(2Q_0)
\\
        Q &= -\tfrac23 Q_0 \e^\tau x \Big[ 3[(w\e^{\tau}\s x)^2+1]
        [ (f_2^2-1)w + 2f_2(w\t-2)\cosh w\tau ]
\notag\\
        &\qquad +4w\e^{2\tau}x^2(w^2+2-3w\t)\Big] \e^{-3\tau}
                + Q_3
\\
        \lambda &=-6\ln(\s) + 4\ln[(w\e^{\tau}\s x)^2+1]+2\ln[f_2^2+1]
                -(w^2+9)\tau + \lambda_3,
\end{align}
where $f_2$ is the factor $Q\e^P$ of the spike solutions.
These second-order spike solutions have multiple zeros of $\Nm$ within the 
horizon of $x=0$. The solutions are parameterized by $w$ and $Q_2/Q_0$.

The variables $(\Sm,\Nc,\Sc,\Nm)$ of the third-order spike solutions can
be obtained. $\Nm$ has even more zeros within the horizon
of $x=0$, These solutions are parametrized by $w$, $Q_2/Q_0$ and
$Q_3/Q_0$.

These second and third-order spike solutions have not been commonly seen
in numerical simulations because initial data with multiple zeros of 
$\Nm$ within one horizon are not commonly chosen in numerical 
simulations.

It is of interest to study the movement of these zeros over more than one 
Kasner era, in the context of $G_2$ models.
If multiple zeros stay within the same horizon, then
high-order spike solutions are conjecture to be part of the 
generalized Mixmaster 
attractor. Conversely, if they eventually move away from each other's 
horizon, then all high-order spike solutions gradually splinter into 
first-order spikes, and are not part of the attractor. 
I shall leave the detailed analysis of higher-order spikes for future 
works, and note its potential relation with the high-velocity spike 
approximation in \cite[Section 6]{RW01}.

\subsection{Other seed solutions}

Seed solutions other than the Kasner and Taub solutions can be used. 
The following are examples of explicit solutions in the class of Gowdy 
spacetimes.
\begin{itemize}
\item	Solutions with $Q=\text{const}$ and separable $P=F(\tau) G(x)$
	and their linear combinations \cite[Section 3]{B74}.
\item   Solutions (apparently new) with $Q=\text{const}$ and additively 
	separable $P$:
	\be
        P = c_1(2x^2+\e^{-2\tau}) + c_2 \tau + c_3 x + c_4.
	\ee
\item   The Wainwright-Marshman solution
	\cite[Case I with $m=-\tfrac{3}{16}$]{WM79}, 
	\cite[Section 6.2]{thesis}.
\end{itemize}
But because the Kasner and Taub solutions are part of the
generalized Mixmaster 
attractor while these other solutions are not, any spike solutions 
generated from these solutions are not expected to be part of the 
attractor.

\section{Conclusion}

Gowdy spacetimes have been appreciated as a simple class of 
inhomogeneous spacetimes for both manageable mathematical and numerical 
analyses. Even so, the discovery of the spike solutions is a pleasant 
surprise.

The main results of this paper are the derivation of the explicit spike 
solutions using a solution-generating transformation, and the conjecture 
that the spike solutions are part of the generalized Mixmaster 
attractor.
Higher-order spikes remain to be analyzed, and may also be part of the 
attractor.

A similar solution-generating method was used in \cite{BM00} to 
generate exact inhomogeneous $U(1)$ symmetric solutions with Mixmaster 
dynamics.
Although these solutions contain $G_2$ solutions as a special case, 
comparison with the spike solutions is difficult due to the difference in 
the temporal gauges used. Nonetheless, it is important to examine if 
these solutions also contain spikes.

It is reassuring to see the spike solutions very closely resemble those 
seen numerically. 
The spike solutions will be immensely valuable to the study of 
spikes in more general classes of spacetimes.
I expect the spike solutions in this paper to be only one of 
several possible kinds of spikes in generic spacetimes.
Insufficiently resolved spiky structures in the singular regime have been 
observed in numerical simulations of $U(1)$ symmetric spacetimes 
\cite{BM98},\cite[Chapter 7]{H99} and generic spacetimes \cite{G04}.
Therefore, the ultimate goal is to understand these spikes, and to
determine whether they are part of the 
generalized Mixmaster attractor in generic spacetimes. 
Numerical study of spikes in $G_2$ cosmologies in collaboration with David 
Garfinkle, Frans Pretorius and Lars Andersson is in progress.

\section*{Acknowledgment}

I thank Lars Andersson for sponsoring my visit at the Albert 
Einstein Institute, during which the discovery was made. In addition to 
Lars, I also thank
Beverly Berger,
Piotr Chru\'{s}ciel,
Jim Isenberg,
Vincent Moncrief,
Frans Pretorius,
Alan Rendall
and
Claes Uggla
for helpful discussions.
Usage of symbolic computation software {\tt MAPLE} and numerical software 
{\tt MATLAB} has been indispensable.

\appendix
\section{The Weyl scalar invariants}

The orthonormal frame components $C_{abcd}$ of the Weyl tensor 
can be conveniently expressed in terms of the electric and magnetic 
components $E_{\alpha\beta}$ and $H_{\alpha\beta}$ \cite{EU97}:
\be
	C_{\alpha0\beta0}=E_{\alpha\beta},\qquad
	C_{\alpha\beta\gamma\delta}=
	-\epsilon^\mu{}_{\alpha}\epsilon^\nu_{\gamma\delta}E_{\mu\nu}
	,\qquad
	C_{\alpha\beta\gamma0}=\epsilon^\mu{}_{\alpha\beta}H_{\gamma\mu},
\ee
which are then normalized by $3\beta^2$:
\be
	\E_{\alpha\beta}=\frac{1}{3\beta^2}E_{\alpha\beta},\qquad
	\H_{\alpha\beta}=\frac{1}{3\beta^2}H_{\alpha\beta},
\ee
and further decomposed the same way as the shear matrix. The components 
are given by
\begin{align}
\Ep &= =\tfrac13\Sp-\tfrac13(\Sm^2+\Sc^2)+\tfrac23(\Nm^2+\Nc^2)
\\
\Em &= \tfrac13(1-3\Sp)\Sm+\tfrac23\Np\Nm+\tfrac13(\di-r)\Nc
\\
\Ec &= \tfrac13(1-3\Sp)\Sc+\tfrac23\Np\Nc-\tfrac13(\di-r)\Nm
\\
\Hp &= -\Nm\Sm-\Nc\Sc
\\
\Hm &= -\Sp\Nm-\tfrac23\Np\Sm-\tfrac13(\di-r)\Sc
\\
\Hc &= -\Sp\Nc-\tfrac23\Np\Sc+\tfrac13(\di-r)\Sm
\end{align}
where $\Np=\sqrt{3}\Nm$, $r=-3(\Nc\Sm-\Nm\Sc)$.
The four Weyl scalar invariants are computed as follows
\begin{align}
	C_{abcd}C^{abcd} 
&= 8 (E_{\alpha\beta}E^{\alpha\beta} - H_{\alpha\beta}H^{\alpha\beta})
\\
	C_{abcd}{}^*C^{abcd}
	&= 16 E_{\alpha\beta} H^{\alpha\beta}
\\
	C_{ab}{}^{cd}C_{cd}{}^{ef}C_{ef}{}^{ab}
	&= -16 (E_\alpha{}^\beta E_\beta{}^\gamma E_\gamma{}^\alpha
	-3 E_\alpha{}^\beta H_\beta{}^\gamma H_\gamma{}^\alpha)
\\
	C_{ab}{}^{cd}C_{cd}{}^{ef} {}^*C_{ef}{}^{ab}
        &= 16 (H_\alpha{}^\beta H_\beta{}^\gamma H_\gamma{}^\alpha
        -3 E_\alpha{}^\beta E_\beta{}^\gamma H_\gamma{}^\alpha),
\end{align}
where ${}^*C_{abcd}=\tfrac12\eta_{ab}{}^{ef}C_{efcd}$,
and $\eta^{abcd}$ is the totally antisymmetric permutation tensor, with 
$\eta^{0123}=1$.



\begin{thebibliography}{99}

\bibitem{BM93}
Berger B K and Moncrief V 1993 
Numerical investigation of cosmological singularities 
\prd {\bf 48} 4676--4687

\bibitem{BG98}
Berger B K and Garfinkle D 1998
Phenomenology of the Gowdy universe on $T^3 \times R$
\prd {\bf 57} 4767--4777

\bibitem{HS98}
Hern S D and Stewart J M 1998
The Gowdy $T^3$ cosmologies revisited
\cqg {\bf 15} 1581--1593

\bibitem{RW01}
Rendall A D and Weaver M 2001 
Manufacture of Gowdy spacetimes with spikes
\cqg {\bf18} 2959--2975

\bibitem{GW03}
Garfinkle D and Weaver M 2003 
High velocity spikes in Gowdy spacetimes
\prd {\bf67} 124009

\bibitem{AELU05}
Andersson L, van Elst H, Lim W C and Uggla C 2005
Asymptotic silence of generic cosmological singularities
\prl {\bf94} 051101

\bibitem{thesis}
Lim W C 2004
Ph.D. thesis, University of Waterloo
arXiv:gr-qc/0410126

\bibitem{LK63}
Lifshitz E M and Khalatnikov I M 1963
Investigation in relativistic cosmology
{\it Adv. Phys.} {\bf 12} 185--249

\bibitem{BKL70}
Belinskii V A, Khalatnikov I M and Lifshitz E M 1970
Oscillatory approach to a singular point in the relativistic cosmology
{\it Adv. Phys.} {\bf 19} 525--573

\bibitem{BKL82}
Belinskii V A, Khalatnikov I M and Lifshitz E M 1982
A general solution of the Einstein equations with a time singularity
{\it Adv. Phys.} {\bf 31} 639--667

\bibitem{M69a}
Misner C W 1969
Mixmaster universe
\prl {\bf 22} 1071--1074

\bibitem{M69b}
Misner C W 1969
Quantum cosmology I 
{\it Phys. Rev.} {\bf 186} 1319--1327

\bibitem{MTW}
Misner C W, Thorne K S and Wheeler J A 1973
{\it Gravitation}
W H Freeman and Sons

\bibitem{UEWE03}
Uggla C, van Elst H, Wainwright J and Ellis G F R 2003
Past attractor in inhomogeneous cosmology
\prd {\bf68} 103502

\bibitem{HUR07}
Heinzle J M, Uggla C and R\"{o}hr N 2007
The cosmological billiard attractor
arXiv:gr-qc/0702141

\bibitem{EUW02}
van Elst H, Uggla C and Wainwright W 2002
Dynamical systems approach to $G_2$ cosmology
\cqg {\bf19} 51--82

\bibitem{H99}
Hern S D 1999
Ph.D. thesis, University of Cambridge
arXiv:gr-qc/0004036

\bibitem{G04}
Garfinkle D 2004
Numerical simulations of generic singularities
\prl {\bf 93} 161101

\bibitem{EU97}
van Elst H and Uggla C 1997
General relativistic 1+3 orthonormal frame approach
\cqg {\bf 14} 2673--2695


\bibitem{WE97}
Wainwright J and Ellis G F R (eds) 1997
{\it Dynamical Systems in Cosmology}
Cambridge University Press, Cambridge


\bibitem{B74}
Berger B K 1974
Quantum graviton creation in a model universe
{\it Ann. Phys.} {\bf83} 458--490

\bibitem{WM79}
Wainwright J and Marshman B J 1979
Some exact cosmological models with gravitational waves
{\it Phys. Lett. A} {\bf72} 275--276

\bibitem{BM00}
Berger B K and Moncrief V 2000
Exact $U(1)$ symmetric cosmologies with local mixmaster dynamics
\prd {\bf 62} 023509

\bibitem{BM98}
Berger B K and Moncrief V 1998
Evidence for an oscillatory singularity in generic U(1) symmetric 
cosmologies on $T^{3}\times— R$
\prd {\bf 58} 064023

\end{thebibliography}
\end{document}